\documentclass[sigconf]{acmart}

\AtBeginDocument{%
  \providecommand\BibTeX{{%
    \normalfont B\kern-0.5em{\scshape i\kern-0.25em b}\kern-0.8em\TeX}}}

\setcopyright{acmcopyright}
\copyrightyear{2024}
\acmYear{2024}
\acmDOI{XXXXXXX.XXXXXXX}

\acmConference[SIGCSE 2024]{Proceedings of the 55th ACM Technical Symposium on Computer Science Education V. 1}{March 20--23, 2024}{Portland, OR, USA} \acmBooktitle{Proceedings of the 55th ACM Technical Symposium on Computer Science Education V. 1 (SIGCSE 2024), March 20--23, 2024, Portland, OR, USA}

\acmPrice{15.00}
\acmISBN{978-1-4503-XXXX-X/18/06}

\usepackage{graphicx}
\usepackage[printonlyused,nohyperlinks]{acronym}

\usepackage{stfloats}
\usepackage{placeins}
\usepackage{multirow}

\begin{document}

\title{EIT: Earnest Insight Toolkit for Evaluating Students’ Earnestness in Interactive Lecture Participation Exercises}

\author{Mihran Miroyan}
\authornote{Authors contributed equally to this research.}
\orcid{0009-0001-6338-6671}
\affiliation{%
  \institution{University of California, Berkeley}
  \city{Berkeley}
  \state{CA}
  \country{USA}
}
\email{miroyan.mihran@berkeley.edu}

\author{Shiny Weng}
\orcid{0009-0008-1614-1277}
\authornotemark[1]
\affiliation{%
  \institution{University of California, Berkeley}
  \city{Berkeley}
  \state{CA}
  \country{USA}
}
\email{shinyweng@berkeley.edu}

\author{Rahul Shah}
\orcid{0000-0002-2262-0409}
\affiliation{%
  \institution{University of California, Berkeley}
  \city{Berkeley}
  \state{CA}
  \country{USA}
}
\email{rsha256@berkeley.edu}

\author{Lisa Yan}
\orcid{0009-0007-2310-3060}
\affiliation{%
  \institution{University of California, Berkeley}
  \city{Berkeley}
  \state{CA}
  \country{USA}
}
\email{yanlisa@eecs.berkeley.edu}

\author{Narges Norouzi}
\orcid{0000-0001-9861-7540}
\affiliation{%
  \institution{University of California, Berkeley}
  \city{Berkeley}
  \state{CA}
  \country{USA}
}
\email{norouzi@berkeley.edu}

%
\renewcommand{\shortauthors}{Anonymous et al.}

\begin{abstract}
    In today's rapidly evolving educational landscape, traditional modes of passive information delivery are giving way to transformative pedagogical approaches that prioritize active student engagement. Within the context of large-scale hybrid classrooms, the challenge lies in fostering meaningful and active interaction between students and course content. This study delves into the significance of measuring students' earnestness during interactive lecture participation exercises. By analyzing students' responses to interactive lecture poll questions, establishing a clear rubric for evaluating earnestness, and conducting a comprehensive assessment, we introduce EIT (Earnest Insight Toolkit), a tool designed to assess students' engagement within interactive lecture participation exercises—particularly in the context of large-scale hybrid classrooms. Through the utilization of EIT, our objective is to equip educators with valuable means of identifying at-risk students for enhancing intervention and support strategies, as well as measuring students' levels of engagement with course content.

\end{abstract}




\maketitle

\section{Introduction}

In recent years, active learning has fostered dynamic, participatory learning experiences and replaced many traditional, passive teaching strategies across multiple disciplines in post-secondary education~\cite{prince2004,freeman2014,theobald2020active}. Active learning practices like peer instruction and clicker quizzes have registered  substantial benefits to student learning in the classroom~\cite{flippedclassroom, designpatterns, quickpolls}. However, there remains a limited understanding of measuring \textit{student earnestness}---defined as active, sincere attempts to complete activities, despite easy access to correct solutions~\cite{yuen2016earnestness}---within university-level Computer Science (CS) classrooms. Will a student's earnest completion of a clicker question during class---an active learning activity---correlate with their increased learning, even if there is no course credit tied to a correct answer? 

The challenge of assessing individual learning is exacerbated when there are hundreds or even thousands of students, as is often the case in fundamental computer science or data science courses at larger institutions. In smaller classrooms, instructors often can directly perceive and quantify earnest student behavior. However, in larger classrooms, the magnitude of students and interactions makes manual detection and measurement of learning impractical. Instead, course learning goals are often assessed automatically. For example, instead of manual grading, many programming assessments are graded with autograders, instead of other manual approaches~\cite{mirhosseini2023your, norouzi2018quantitative}, which allows instructors to reflect on the efficacy  and quality of course assignments. However, many instructors cannot directly assess the learning benefits of ``participation-based activities'' where students are graded not on correctness, but rather whether they attempt. However, active learning lecture activities, due to their constructivist nature, are often graded only on participation; as a result, instructors cannot examine student earnestness and the dynamics of individual student learning.

In this paper, we present the Earnest Insight Toolkit (EIT), a semi-automated tool for assessing student earnestness in interactive lecture participation exercises. EIT offers a valuable tool for CS educators to identify students who are less engaged with the course and are at risk of dropping the class, thus enhancing interventions and support strategies \cite{atriskstudents}. We evaluate this tool in the context of a large\footnote{On the order of 1000 students; precise classroom size is redacted for peer review.} data science fundamentals classroom with multimodal, hybrid lecture instruction (synchronous/asynchronous and in-person/online). We also propose a metric for differentiating correctness and earnestness in student responses. Through the use of EIT, educators of large computing classrooms can glean more valuable insights into the efficacy of their teaching through active learning exercises.


\section{Related Work}

\textit{Hybrid instruction}. In the past few years, the education landscape has experienced a notable shift attributed to technological progress, evolving pedagogical approaches, and the disruption caused by the pandemic~\cite{ng2021timely}. The concept of hybrid instructional models has gained considerable traction as educators strive to integrate the strengths of conventional in-person teaching with the capabilities of digital learning platforms, all while responding to external influences~\cite{triyason2020hybrid}. The year 2020 marked a significant transition to online education on a global scale when the widespread outbreak of COVID-19 necessitated the suspension of traditional in-person teaching across the globe. 

To understand the shift to online instruction, Hofer et al. discussed the broad framework of Communities of Practice (CoPs) combined with the integrative framework for learning activities involving technology in higher education. They shared insights into online learning, emphasizing emergency online teaching and learning scenarios~\cite{hofer2021online}. Researchers have also investigated interventions for effective online instruction. Researchers discuss the effectiveness of group work activities in synchronous online classroom spaces as an example of an active learning environment~\cite{raes2020systematic, lakhal2021students}.

\textit{Large Classrooms.}
In higher education, addressing the unique challenges of large classroom settings has been a longstanding concern. While efficient in disseminating information, traditional lecture-based instruction often struggles to engage diverse and sizable student populations effectively and, in many cases, results in equity gaps~\cite{norouzi2023equity}. Researchers have explored various pedagogical strategies to enhance learning experiences in large classrooms~\cite{tracy2020successful}. Active learning techniques have emerged as a prominent avenue to foster greater student engagement, participation, and interaction~\cite{theobald2020active}. These approaches emphasize collaborative activities, discussions, problem-solving, and peer engagement, shifting the focus from a passive learning environment to one characterized by dynamic participation. The need to improve educational outcomes within large classrooms has led to exploring innovative methodologies that utilize technology, spatial design, and teaching practices. As institutions grapple with the implications of evolving instructional paradigms, a deeper understanding of active learning's impact on large classrooms becomes increasingly critical.

\textit{Active Learning.}
Active learning strategies, such as encouraging students to answer classroom questions, have emerged as tools for transforming students' learning experiences in large-scale classrooms. Active learning fosters engagement, deepens understanding, and promotes a more inclusive educational environment~\cite{sitthiworachart2022technology}. Moreover, active learning strategies offer meaningful insights into student progress, helping educators identify and support those at risk.

Researchers reported on using technology to enhance active learning techniques within large-scale classrooms~\cite{martyn2007clickers}. They also advocate using mobile phones instead of laptops and specialized wireless devices such as clickers~\cite{litchfield2007directions}. Köppe et al. analyze various patterns for introducing activities into lectures to engage students and encourage active learning~\cite{designpatterns}. More concretely, the authors suggest that question formation is critical to sparking students' involvement and participation. Instead of merely asking questions, planning and creating questions in advance that actively engage students and align with the lesson's goals is more effective. This may include simple, straightforward, and close-ended questions, drawing students into the lecture, as they are more likely to know the answers.
Additionally, beyond simple questions, open-ended questions require students to provide thoughtful answers based on their existing knowledge. It has also been shown that when students are asked about the reasoning behind their answers, it stimulates deeper thinking and encourages thoughtful responses rather than random guesses. In this work, we report on our experience in using the Slido Application for live polling during the lecture synchronously while providing students the option to respond to learning questions asynchronously as they watch the lecture recordings~\cite{slido}.

\section{Data Collection}

\subsection{Classroom Setup}

We used student responses to interactive lecture participation exercises conducted within a large\footnotemark[\value{footnote}] Data Science class at a large public university in the United States during a 14-week semester in Spring 2023. Instructors of the course utilized Slido, a Q\&A and polling platform for live and virtual meetings and events, to actively engage students by incorporating live polls throughout the lecture.

The classroom setup and lecture participation exercises include synchronous and asynchronous lecture participation options in a hybrid classroom environment. 
For each lecture, instructors presented up to three questions during the lecture; the types and format of questions are discussed in the following subsection. Students could see the live results on the projected screen while responding to questions; after a few minutes, the instructor would reveal the correct answer, or in the case of open-ended questions, they discussed potential correct approaches to the question. Lecture attendance corresponded to the 5\% of their final grade. Responses were graded on completion rather than correctness, and students could miss up to 3 lectures and receive full attendance.  Students had two options to get credit for lecture attendance:

1. \textbf{Synchronous participation}: The lectures were broadcast live through Zoom. Thus, students were given the opportunity to get credit for both in-person and online synchronous participation. For synchronous participation, students must answer at least one out of all the questions to get full credit for lecture attendance.

2. \textbf{Asynchronous participation}: As the recordings of the lectures were provided to students, students who could not get credit for synchronous participation were given a week after the lecture to get credit for asynchronous participation. However, for asynchronous participation, students must answer all three questions for full participation credit. As for synchronous participation, the responses were graded on completion.

\subsection{Lecture Poll Data}
Collected poll data over 28 lectures yielded 84 total questions, categorized into multiple-choice and word cloud. Multiple-choice questions (65 questions; 24,730 responses) had between 2 and 8 choices with correct answer(s).  Word cloud questions (19 questions; 9,556 responses) are free-response questions for which student responses are visually projected live in a word cloud format to the class. We seek to avoid conflating earnestness with correctness; because it can be challenging to do so with multiple-choice questions, in this experience report, we only focus on word cloud questions and report on how they can be used to measure students' earnestness.


\begin{table*}[ht]
\caption{A sample of word cloud poll questions associated with each lecture and the number of responses per poll.}
\vspace*{-1ex}
\begin{tabular}{lllrrrr}
\toprule
    &   \multirow{ 2}{*}{Question} &  \multirow{ 2}{*}{Category} & \multirow{ 2}{*}{Lecture \#}  &   Unique &  \multirow{ 2}{*}{\% Synchronous} &  \multirow{ 2}{*}{Total samples} \\
    &   &  &  & samples \\
\midrule
$Q_1$  &  Why do you think the data science lifecycle is iterative? &  Reflection &        1 &               654 &          60.56 &             1288 \\
\hline
$Q_2$  &  Write a groupby.agg call that returns the total number of & & & & & \\
&
 babies born every year.
The image here shows the & & & & & \\
&
example of calculating ratio\_to\_peak. &      Coding &        3 &               346 &          56.93 &              808 \\
\hline
$Q_3$  &  Write a groupby.agg call that returns the total number  & & & & & \\
&
of babies with each name. 
The image here shows the  & & & & & \\
&
example of calculating ratio\_to\_peak. &      Coding &        3 &               430 &          68.01 &             1088 \\
\hline
$Q_4$  &  How many observations are in the bin [110, 115) given the & & & & & \\
& following information? 
There are 1174 observations in total. & Numerical &        7 &               188 &          54.05 &             1025 \\
\hline
$Q_5$  &  Why don't we use residual error directly and & & & & & \\
& instead we use absolute loss or squared loss? &  Conceptual &       10 &               723 &          46.33 &             1049 \\
\hline
$Q_6$  &  Suppose we pick k = 3 and we have 4 possible   & & & & & \\
& hyperparameter values alpha=[0.01, 0.1, 1, 10]. How  & & & & & \\
& many total MSE values will we compute to get the & & & & & \\
&  quality of alpha=10? &   Numerical &       15 &                80 &          39.07 &             1070 \\
\hline
$Q_7$  &  Why did we shuffle the data before selecting the training  & & & & & \\
& and validation sets? &  Conceptual &       15 &               595 &          42.78 &             1143 \\
\hline
$Q_8$  &  Let X be the outcome of a single fair die roll. What is the & & & & & \\
& variance of $X$ ($\text{Var}[X]$)? &   Numerical &       16 &               170 &          20.16 &              997 \\
\hline
$Q_9$  &  Let X be the outcome of a single fair die roll. What is the & & & & & \\
& expectation of $X$ ($E[X]$)? &   Numerical &       16 &               123 &          31.50 &             1165 \\
\hline
$Q_{10}$  &      Write down one take-away from today's lecture. &  Reflection &       18 &               768 &          27.94 &             1643 \\
\hline
$Q_{11}$ &  What is the difference between model risk  &   &        &                &           &               \\
&  and empirical risk? & Conceptual &       20 &               661 &          43.46 &              803 \\
\hline
$Q_{12}$ &  What function maps a positive real number to an  & & & & & \\
& unbounded real number? &  Conceptual &       23 &               225 &          43.63 &             1036 \\
\hline
$Q_{13}$ &  We have 100 emails, of which only 5 are truly spam, and  & & & & & \\
& the remaining 95 are ham. Assume that we have a model   & & & & & \\
& that classifies all emails as ham. & & & & & \\
& What is the accuracy of the classifier? &   Numerical &       24 &               104 &          23.75 &              800 \\
\hline
$Q_{14}$ &  Why would one perform PCA before training a Logistic & & & & & \\
& Regression model? &  Conceptual &       26 &               319 &          31.52 &              552 \\
\hline
$Q_{15}$ &  What is the highest value of S, given the information  & & & & & \\
&in the image? &   Numerical &       27 &                28 &          95.44 &              241 \\
\hline
$Q_{16}$ &  What are some good evaluation metrics to compare  & & & & & \\
& different clustering results? &  Conceptual &       27 &                99 &          96.26 &              294 \\
\bottomrule
\end{tabular}
\label{tab:listofquestions}
\end{table*}


Table~\ref{tab:listofquestions} lists a sample of word cloud questions along with categories, lecture number, number of unique responses, percentage of synchronous responses, and the total number of responses per question.
We report on four main categories of word cloud questions that instructors utilized: open-ended reflection questions, close-ended conceptual questions, coding questions, and numerical questions. Open-ended reflection questions (4 questions, 19.5\% of responses) tend to relate to Data Science topics, but there may be many different responses with no single correct answer. An example of the open-ended general question is $Q_1$,``Why do you think the Data Science life cycle is iterative?'' Close-ended conceptual questions (6 questions, 32.5\% of responses) are closely tied to the lecture content and have few correct solutions, such as $Q_{11}$, ``What is the difference between model risk and empirical risk?'' Coding questions (2 questions, 12.6\% of responses) refer to topics such as Pandas and regular expression. Finally, numerical questions (6 questions, 35.3\% responses) are mathematical, such as $Q_{13}$, ``We have 100 emails, of which only 5 are truly spam, and the remaining 95 are ham. Assume that we have a model that classifies all emails as ham. What is the accuracy of the classifier?''

On average, 47.4\% of word cloud responses were synchronous. Synchronous participation was particularly low during exam periods ($Q_8$ and $Q_9$: Lecture 16, was on the same day as the course midterm), implying many students did not attend and opted for asynchronous catch-up. Overall participation at the end of the semester ($Q_{15}$ and $Q_{16}$: Lecture 27) was especially low, perhaps because many students had received full attendance already and decided against completing the lecture questions.

\textit{Browser participation software}: The data is collected from Slido, a browser-based participation meeting tool; in our case, students used Slido to submit responses to questions synchronously or asynchronously through their personal devices. EIT framework is generalizable to other active learning participation technologies such as iClicker devices~\cite{iClicker}, pollEverywhere~\cite{sellar2011poll}, and Google Forms. The application is freely available to all students.

\section{Methodology}


Following a comprehensive review of the Slido Word Cloud questions featured in the lectures, we used Table~\ref{tab:listofquestions} to select five questions ($Q_1$, $Q_5$, $Q_7$, $Q_{11}$, and $Q_{14}$) that we believe can be effectively used to measure students' earnestness.



To enhance the precision of this assessment, we devise a rubric rating the earnestness of each response on a scale of 1 to 5. We aim to evaluate the level of effort students invested in their responses, irrespective of correctness. The rubric is provided in Table~\ref{tab:rubric}. 

\begin{table}[tb]
  \caption{Rubric for Measuring Earnestness in Word Cloud Responses}
  \label{tab:rubric}
  \begin{tabular}{cl}
    \toprule
    Rubric&Score\\
    \midrule
    1 & Completely no effort demonstrated (e.g., blank, \\
      & emoji,  random characters). \\
    2 & Potential effort, but unclear how the response is \\
      & connected to the question. \\
    3 & Minimal effort related to the question, but is vague. \\
    4 & Some effort and reasoning are attempted. \\
    5 & Considerable effort and explicitly displays reasoning. \\
  \bottomrule
\end{tabular}
\end{table} 

To better interpret the devised rubric to support the binary classification task, we classify responses earning values of 4 and 5 as earnest and responses earning values of 1 or 2 as non-earnest. The responses rated as 3 are not used in the classification task.

\subsection{Data Labeling and Sampling}

We performed a procedural data cleaning, which involved lower-casing responses and addressing any formatting inconsistencies. Next, we randomly sampled student responses to Word Cloud questions for testing purposes. To account for the class imbalance between earnest and non-earnest responses, for each of the five selected questions, we first found the following metrics:

\begin{enumerate}
\item Euclidean distance between the embedded response and the center of the unique responses for the question using Roberta~\cite{liu2019roberta} sentence transformer model. We assume that the distance from the center is inversely proportional to the earnestness of the response. 
\item Frequency of the response. Based on our initial data exploration, we observe that popular responses tend to be more earnest. 
\item Edit distance from the most common response. This metric accounts for any typos and can be particularly useful for analyzing coding questions.
\item Number of characters in the response text. This metric is essential since we observed that some students put in random characters in contrast with students who put considerable effort and answered the questions in 1-2 thorough sentences. Thus, we assume that the length of the response is positively correlated with earnestness.
\end{enumerate}

As we observed a significant class imbalance---far more earnest responses---we used a rule-based sampling procedure to sample approximately an equal number of earnest and non-earnest responses. We give the 20\% of the data in the direction of the non-earnest extreme (depending on the metric we use) an equal chance of being sampled as the rest of the data. We first sample around 20\% of unique responses using the rule-based sampling procedure described above for each proposed metric, then combine unique entries across each sample and then sample 200 responses from the combined set. Thus, we obtain 200 unique responses for each question (with the exception of one question that had less than 200 unique responses), accounting for the class imbalance in the original dataset.

After obtaining random samples for each question, we proceeded to manually annotate the responses based on the devised rubric (Table~\ref{tab:rubric}). As earnestness can be subjective, three individuals labeled each response manually, and the final label was determined by averaging the scores provided by the three annotators. The inter-rater reliability score was 70.6\% for 3-class labeling: non-earnest (1 and 2), neutral (3), and earnest (4 and 5).

\subsection{Earnest Insight Toolkit}

In this paper, we propose EIT (Earnest Insight Toolkit) as a tool for identifying non-earnest responses and tracking students' earnest performance across the semester. To identify non-earnest responses, we initially employ a supervised K-Nearest Neighbors (KNN) model. For each of the five Word Cloud questions, we calculate the embeddings of the responses using Roberta~\cite{liu2019roberta}, a sentence transformer model capable of mapping text to a 768-dimensional vector space. Following the embedding calculation, we utilize t-distributed stochastic neighbor embedding (t-SNE) to project these high-dimensional embeddings into a 2-dimensional space. 

\begin{figure*}[t]
     \centering
     \begin{tabular}{cc}
\raisebox{-\totalheight}{\includegraphics[width=0.36\textwidth]{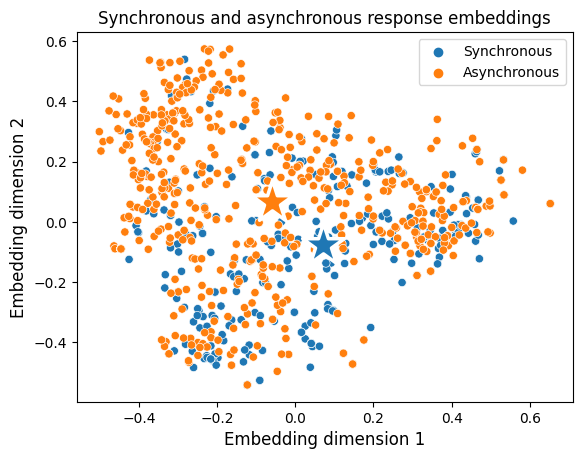}}
      & 
      \raisebox{-\totalheight}{\includegraphics[width=0.36\textwidth]{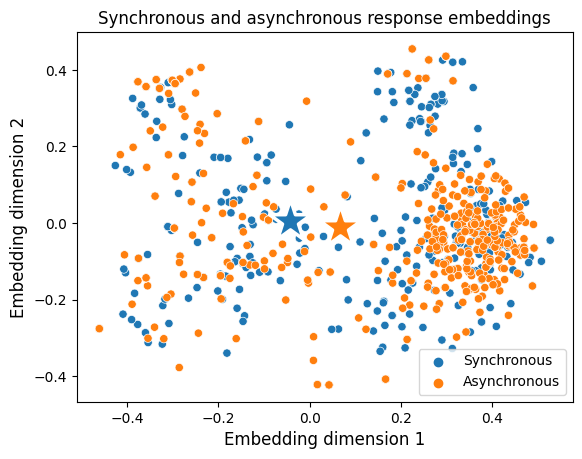}}
 \end{tabular}
      \begin{flushleft}
        
      \end{flushleft}
     \caption{
      \label{fig:sync_embeddings} Synchronous vs. Asynchronous Response Embeddings. (\textbf{Left}):  Why don't we use residual error directly and instead we use absolute error or squared loss? (\textbf{Right}): Why do you think the data science lifecycle is iterative?}

\end{figure*}

Additionally, after computing the embeddings of student responses from synchronous and asynchronous participation, it became evident that the difference between the embeddings corresponding to the responses from the two groups was negligible (refer to Figure~\ref{fig:sync_embeddings}). Consequently, for the main analysis and toolkit development, we combined the responses from both asynchronous and synchronous groups.

\begin{figure*}[tb]
     \centering
     \begin{tabular}{cc}
     
     \raisebox{-\totalheight}{\includegraphics[width=0.35\textwidth]{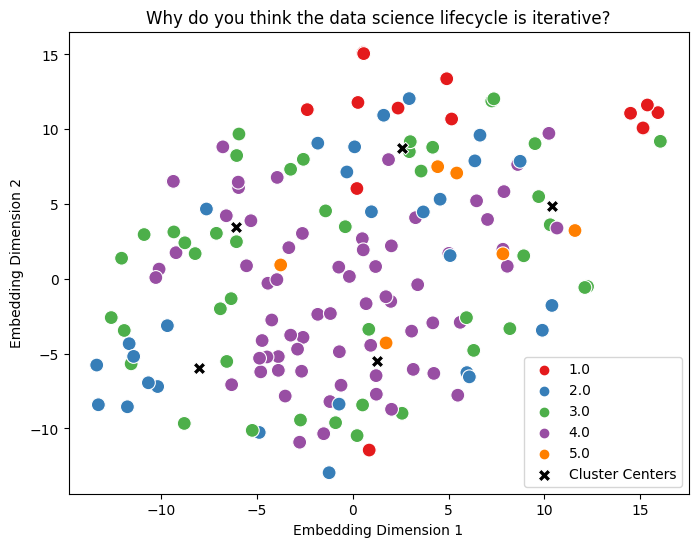}}
      & 
      \raisebox{-\totalheight}{\includegraphics[width=0.35\textwidth]{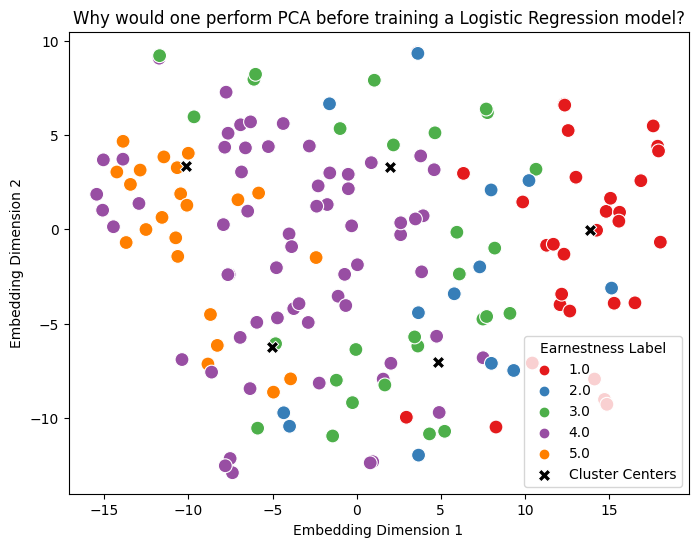}}  \\ 
\raisebox{-\totalheight}{\includegraphics[width=0.35\textwidth]{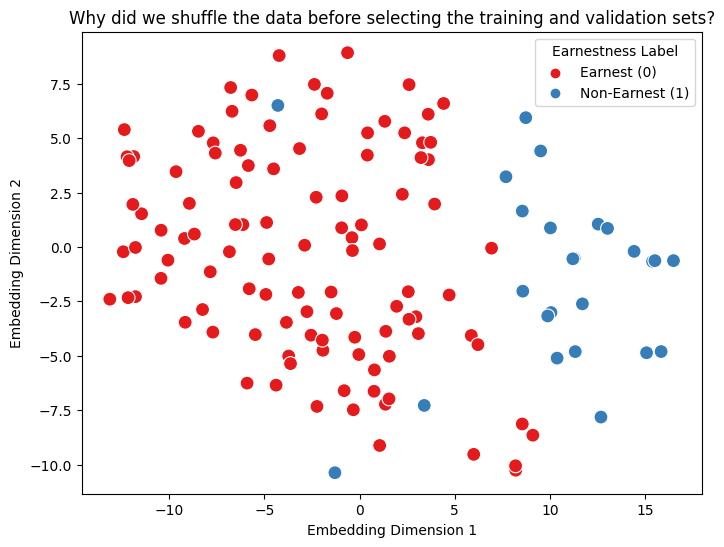}}
      &  
      \raisebox{-\totalheight}{\includegraphics[width=0.35\textwidth]{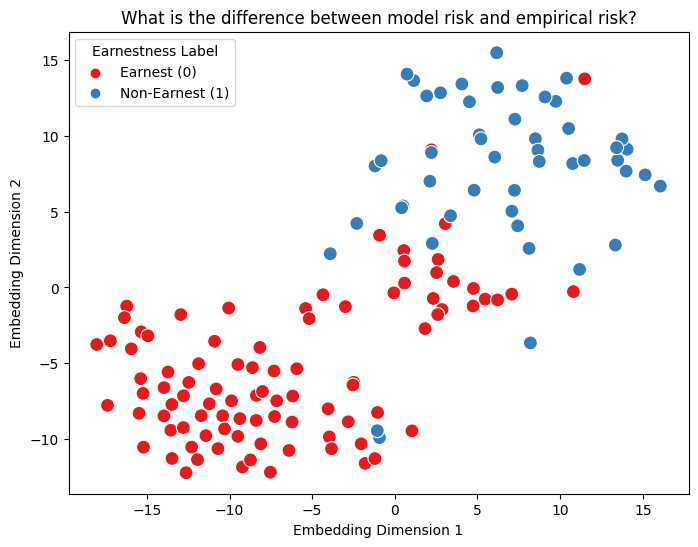}} 
      \vspace*{-2ex}
      \end{tabular}
     \caption{
     \label{fig:clusters}Earnestness clustering results for different word cloud questions using multi-class and binary labels.}

\end{figure*}

Figure~\ref{fig:clusters} shows the projected embeddings of unique responses. The clusters of neutrality (earnestness value of 3) do not show a distinct separation from earnest and non-earnest clusters. However, non-earnest categories (rated as 1 or 2) and earnest categories (rated as 4 or 5) demonstrate a clear separability. In alignment with our earnestness rubric, we opted to classify values below three as non-earnest and values exceeding three as earnest. This classification approach and our objective of identifying at-risk students prompted our focus on a binary classification task rather than multi-class classification. We can effectively identify and distinguish clusters associated with non-earnestness and earnestness through binary classification.

Since the main objective of EIT is to detect non-earnest responses for identifying at-risk students, we optimize our classifier for recall. Furthermore, EIT does not require extensive manual labeling of student responses. Specifically, we develop a dataset of non-earnest responses; it combines non-earnest responses for the five labeled questions.
Since non-earnest responses are unrelated to the original question, they are highly likely to also be non-earnest for other sets of questions. Then, we inputted a limited set of labeled earnest responses, which were easily identified through the most frequent responses. Finally, as our initial training set, we ran experiments with different proportions of earnest and non-earnest labeled responses to show the framework's performance under different percentages of labeled samples (refer to Table~\ref{tab:ablation}).

\section{Results}

Based on the plots in Figure~\ref{fig:clusters}, we observe that specific questions lead to the formation of more distinct clusters. For example, questions with readily detectable clusters, like $Q_5$, (``Why don't we use residual error directly and instead we use absolute loss or squared loss?'') are typically less open-ended and will also be relevant to the provided lecture. Questions such as $Q_1$ (``Why do you think the data science lifecycle is iterative?'') are more open-ended. Therefore, we conclude that posing questions to students closely aligned with the lecture content enables a more effective differentiation between non-earnest and earnest responses in large-scale classrooms. 

We conduct KNN 5-fold cross-validation to predict earnestness, utilizing our rubric (ranging from 1 to 5); the results for the five questions are presented in Table~\ref{tab:accuracy}. EIT achieves recall scores of above 0.9 across all listed questions, hence demonstrating its effectiveness in determining non-earnest responses. In consideration of the various question types, such as open-ended versus close-ended questions, EIT's recall score for both types of questions is comparable. Despite the lower accuracy scores in open-ended questions, recall scores for both open and close-ended questions were within a similar range.

\begin{table}[tb]
  \caption{
\label{tab:accuracy}EIT's 5-fold KNN classification accuracy and recall for different types of questions.}
\vspace*{-1ex}
  \begin{tabular}{clll}
    \toprule
    Question&Accuracy&Recall\\
    \midrule
    $Q_1$ & 0.83 & 0.97 \\
    $Q_5$ & 0.98 & 1.00 \\

    $Q_7$ & 0.98 & 0.92 \\
    
    $Q_{11}$ & 0.94 & 0.99 \\
    
    $Q_{14}$ & 0.83 & 1.00 \\
  \bottomrule
\end{tabular}
\end{table} 

\begin{table}[tb]
  \caption{\label{tab:ablation}Experimental results with varying percentages of labeled samples.}
\vspace*{-1ex}
  \begin{tabular}{c|c|c|c}
    \toprule
    Non-earnest (\%) & Earnest (number) & Accuracy & Recall\\
    \midrule
    10\% & 5 & 0.611211 & 0.857039 \\
    10\% & 10 & 0.710311 & 0.765152 \\
    10\% & 20 & 0.825400 & 0.675483 \\
    25\% & 5 & 0.525843	 & 0.911765 \\
    25\% & 10 & 0.635636 & 0.837462 \\
    25\% & 20 & 0.763745 & 0.777548 \\
    50\% & 5 & 0.410610 & 0.972222 \\
    50\% & 10 & 0.551624 & 0.913080 \\
    50\% & 20 & \textbf{0.713383} & \textbf{0.881334}
    \\
\end{tabular}
\end{table}

Table~\ref{tab:ablation} outlines the EIT's accuracy and recall when utilizing different percentages of labeled earnest and non-earnest responses across the five label question responses. Results demonstrate that the framework achieves high recall in identifying non-earnest responses while only utilizing as low as 5 earnest cases (which can easily be provided to the model based on the most frequent responses without extensive manual labeling). However, it is worth mentioning that there is a trade-off between high recall and the accuracy of our classifier. Based on Table~\ref{tab:ablation}, for high recall values, we observe low accuracy scores due to false positives: earnest responses falsely classified as non-earnest.

Consequently, to avoid too many false positives, we can input 50\% of the samples from the non-earnest response dataset and just 20 earnest responses to the question, which can be found by taking the most frequent responses.

As presented in this paper, through the use of labeled and unlabeled Word Cloud responses, EIT is capable of achieving recall values of over 0.9. With labeled responses, EIT is simultaneously capable of reaching accuracy values of over 0.9. Although increasing the number of labeled responses typically results in improved accuracy and recall, we demonstrate that EIT can still attain relatively high recall values of over 0.9 even with limited number of labeled points. This capability allows us to better identify non-earnest responses and students at risk of falling behind in the course. 

\section{Limitations and Improvements}

While EIT is effective in determining non-earnest responses, it has a few limitations. In particular, of all the Slido responses provided, we were only limited to a set of questions that would align with our rubric for earnestness. More concretely, other questions, such as multiple choice or coding questions, were difficult to classify as non-earnest or earnest, even from a human perspective. Moreover, we aimed to capture earnestness rather than correctness, which may be difficult to extract from multiple-choice, numerical, or coding questions. Another area of improvement is to include response time analysis into EIT, as students may tend to input random characters or wait until the instructor reveals the answers and then respond to the poll.


\section{Conclusion}
In this study, we delved into the importance of assessing students' earnestness throughout interactive lecture participation exercises. We also introduced the Earnest Insight Toolkit (EIT), a specifically crafted tool designed to evaluate students' engagement, especially within the realm of large-scale hybrid classrooms. EIT does not require extensive manual labeling of student responses and remains effective amid a limited set of labels. By examining the impact of active learning, the challenges posed by sizable classrooms, the implications of hybrid instruction, and leveraging students' asynchronous and synchronous Slido responses, our objective was to develop a tool with the ability to identify at-risk students, establish a clear rubric for evaluating students' earnestness, and highlight lecture patterns and student learning styles. Notably, we analyzed how the utilization of active learning strategies, such as interactive poll questions and carefully curated questions during lectures, can dramatically improve students' learning experiences and equip educators with enhanced tools for student intervention and support strategies.

\FloatBarrier
\bibliographystyle{ACM-Reference-Format}
\bibliography{reference}

\end{document}